\documentstyle[twoside,fleqn,epsfig,espcrc2,amssymb]{article}%
\begin{document}
\renewcommand{\refname}{\normalsize\bf References}
\newcommand{\figwidth}{0.9\columnwidth}
\newcommand{\figwidtha}{0.95\columnwidth}

\title{%
  Disorder and Two-Particle Interaction in Low-Dimensional Quantum
  Systems}

\author{%
Rudolf A.\ R\"{o}mer, Michael Schreiber and Thomas Vojta%
\address{Institut f\"{u}r Physik,
                 Technische Universit\"{a}t,
                 09107 Chemnitz, Germany}%
\thanks{This work has been supported by the
Deutsche Forschungsgemeinschaft/SFB 393.}%
}
%
%
\begin{abstract}
\hrule
\mbox{}\\[-0.2cm]

\noindent{\bf Abstract}\\
We review some of the recent results on two-interacting particles
(TIP) in low-dimensional disordered quantum models. Special attention
is given to the mapping of the problem onto random band matrices.  In
particular, we construct two simple, seemingly closely related examples
for which an analogous mapping leads to incorrect results. We briefly
discuss possible reasons for this discrepancy based on the physical
differences between the TIP problem and our examples.
 \\[0.2cm] {\em PACS}: 71.55.Jv,
72.15.Rn, 71.30.+h\\[0.1cm] {\em Keywords}: disordered systems,
electron-electron interactions, TIP.\\ \hrule
\end{abstract}

\maketitle

\section{Introduction}
\label{sec-tip-introduction}

Until 1994, the theoretical and experimental research on transport in
disordered systems clearly supported the scaling hypothesis of
localization for non-interacting electrons \cite{AbrALR79,KraM93}. The
systems studied usually fell into the predicted universality classes,
and, if they didn't, then they could be shown not to be generic
\cite{KraM93}.  However, real electrons of course interact
\cite{Cou1736}, and their interaction is of relevance for the
transport properties of disordered systems \cite{EfrS75,Mot90},
especially in 2D and 1D where screening \cite{AshM76} is less
efficient than in 3D.  The influence of {\em weak} interactions has
been investigated extensively using perturbation theory and the
perturbative renormalization group (RG) \cite{Fin83,BelK94}.  One of
the key results is that the lower critical dimension of the MIT is
$d_c^-=2$ as it is for non-interacting electrons.  The application of
the perturbative RG in 1D \cite{GiaS87,GiaS88,GiaS95,KanF92,Sha90}
has lead to the prediction that all thermodynamic states remain
localized in the presence of repulsive many-body interactions.

Due to the persistent current problem
\cite{AltGI91,ButIL83,ChaWBK91,LevDDB90,MaiCB93,SchSSS98} and new
experiments on 2D electron systems \cite{KraKFP94,KraMBF95,KraSSM96}
which show striking signatures of a metal-insulator transition, these
theoretical considerations received a lot of renewed attention. In
order to theoretically study the effects of the interplay between
disorder and interactions, one should in principle solve a problem
with an exponentially growing number of states in the Hilbert space
with increasing system size. At present, this can be achieved only for
a few particles in 1D \cite{Whi93,Whi98,SchJWP98,SchEV99} and very few
particles in 2D \cite{She99,SonS99,VojES98a,EppSV98,EppVS99}.
However, in 1994 Shepelyansky \cite{She94,She96a} proposed to simply
look at {\em two} interacting particles (TIP) in a random environment.
In particular, he suggested that the two particles would form pairs
even for repulsive interactions such that the TIP pairs would have a
larger localization length than the two single particles (SP)
separately.  Thus the interaction would lead to an enhanced
possibility of transport \cite{Imr96}. The perhaps even more
surprising part of the prediction is that the TIP pairs will have a
localization length $\lambda_2$ such that at pair energy $E=0$
\begin{equation}
  \lambda_2 \propto U^2
  {\lambda_1}^2,
\label{eq-lambda2}
\end{equation}
where $U$ represents the onsite interaction strength and $\lambda_1$
is the SP localization length. Since $\lambda_1 \propto 105/W^2$
\cite{Eco90,KapW81,Tho79,CzyKM81,Pic86} in 1D, this implies large
values of $\lambda_2$ for small disorders $W$. 

The first numerical studies devoted to the TIP problem used the
transfer-matrix method (TMM) \cite{KraM93} to investigate the proposed
enhancement of the pair localization length $\lambda_2$
\cite{She94,FraMPW95}. The TMM of \cite{She94} contained an additional
artificial infinitely-long-ranged interaction that tends to mask the
onsite interaction \cite{RomS98}. The TMM of \cite{FraMPW95} avoids
this problem, but is restricted to small system sizes and results for
localization lengths of $\lambda_2 \approx 300$ had been deduced on
systems of size $M=100$. Therefore, two of us studied the TIP problem
by a different TMM \cite{RomS97a} at large system size $M \gtrsim 300$
and found that (i) the enhancement $\lambda_2/\lambda_1$ decreases
with increasing $M$, (ii) the behavior of $\lambda_2$ for $U=0$ is
equal to $\lambda_1$ in the limit $M\rightarrow\infty$ only, and (iii)
for $U\neq 0$ the enhancement $\lambda_2/\lambda_1$ also vanishes
completely in this limit.  Consequently, we concluded \cite{RomS97a}
that the TMM applied to the TIP problem in 1D measures an enhancement
of the localisation length which is due to the finiteness of the
systems considered. Although Ref.\ \cite{RomS97a} has been criticized
\cite{FraMPW97,RomS97b}, we emphasize that subsequent publications
have shown \cite{Hal96T,HalMK98,HalZMK98} that there are no variants
of TMM that reproduce Eq.\ (\ref{eq-lambda2}). Furthermore, in a later
numerical approach \cite{SonO98}, based on Green function methods,
Song and v.\ Oppen argue that our extrapolations for $M\rightarrow
\infty$ were off by $\approx 11\%$ only, whereas the original TMM of
\cite{FraMPW95} deviated by about a factor of $3$ \cite{SonO98}. Thus
while our criticized TMM results are valid, various other numerical
investigations by other groups
\cite{SonO98,WeiP96,SonK97,AkkP97,BriGKT98,HalKK98,WaiP98,WaiWP99,HalK99,Fra99}
as well as ourselves \cite{LeaRS98,LeaRS99,RomLS99,RomLS99b}
convincingly demonstrated some enhancement. The reason for the failure
of the TMM approach of \cite{FraMPW95,RomS97a} has been explained by
Song and v.\ Oppen \cite{SonO98} by arguing that the TMM measures a
localization length $\lambda_f < \lambda_2$ due to the cigar-shape
geometry \cite{WeiMPF95} of the TIP states.

Reliable numerical approaches to the TIP problem are nowadays based on
the computation of the decay of the Green function
\cite{SonO98,SonK97,Fra99,LeaRS99,OppWM96}.  Other direct numerical
approaches to the TIP problem have been based on the time evolution of
wave packets \cite{She94,BriGKT98,HalKK98,HalK99}, exact
diagonalization \cite{WeiMPF95}, variants of level statistics
\cite{WeiP96,AkkP97} and analysis of multifractal properties
\cite{WaiP98,WaiWP99}, perturbative methods \cite{JacS95,JacSS97} and
mappings to effective models
\cite{FraM95,FraMP96,FraMP97,Imr95,PonS97}.  In these investigations
an enhancement of $\lambda_2$ compared to $\lambda_1$ has been found
as remarked above but the quantitative results tend to differ both
from the analytical prediction in Eq.\ (\ref{eq-lambda2}), and, albeit
less, from each other.  Furthermore, a check of the functional
dependence of $\lambda_2$ on $\lambda_1$ is numerically very expensive
since it requires very large system sizes $M \gg \lambda_2 \gg
\lambda_1$.  Extensions of the original arguments have been proposed
for TIP in 2D \cite{She99,Imr96,RomLS99b,Imr95,OrtC99,BorS97} and 3D
\cite{LagS99}, for TIP close to a Fermi sea \cite{OppW95}, and for
long-range interactions in 1D \cite{RomS98,RomS97a,BriGKT98}.

\begin{figure}
\centerline{\epsfig{file=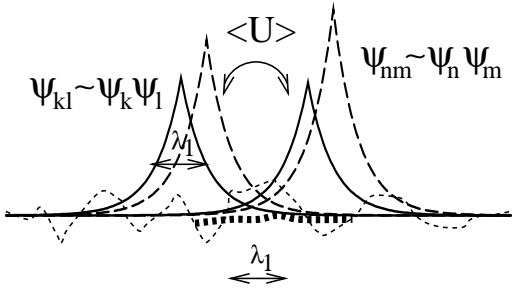,width=\figwidth}}
\vspace{-1cm}
\caption{\label{fig-TIP}
  Schematic picture of the TIP arguments of Ref.\ 
  \protect\cite{She94}.  The two-particle state $\psi_{kl}$ (left
  solid and dashed exponentials indicate the envelopes of the
  constituents $\psi_{k}$ and $\psi{l}$) is localized within a
  distance $\lambda_1$ from the two-particle state $\psi_{nm}$ (right
  solid and dashed curves). The resulting overlap-matrix element
  $u=\langle U \rangle \equiv \langle \psi_{kl} |U| \psi_{nm} \rangle$
  leads to a longer decay length $\lambda_2$ for the TIP state as
  explained in the text.  This effect can be visualized as an
  effective reduction (thick short-dashed line) of the original
  disorder potential (thin short-dashed line).}
\end{figure}

The basic idea leading to the prediction (\ref{eq-lambda2}) is based
on looking at the interaction matrix element between two eigenstates
$\psi_{kl}=\psi_k\psi_l$ and $\psi_{nm}=\psi_n\psi_m$ of the
non-interacting system \cite{She94,She96a}. Here $\psi_k,
\psi_l,\psi_n,\psi_m$ denote SP eigenstates localized with
localization length $\lambda_1$ around sites $k,l,n,m$.  For an onsite
interaction \cite{RomP95} $U \sum_{j=1}^N n_{j\downarrow}
n_{j\uparrow}$ (with $n_{j\tau}$ denoting the number operator at site
$j$ for spin $\tau$) only states with $|k-l|\leq \lambda_1$,
$|n-m|\leq \lambda_1$, $|k-n|\leq \lambda_1$, $|l-m|\leq \lambda_1$
will give significant contributions to the interaction matrix element
\begin{eqnarray}
u & = &\langle \psi_{kl} \vert U \vert \psi_{nm} \rangle \nonumber \\
  & = & U \sum_{j=1}^{N}
           \psi_k^\dagger(j)\psi_l^\dagger(j) \psi_n(j) \psi_m(j).
\end{eqnarray}
These conditions are illustrated in Fig.\ \ref{fig-TIP}.
If one assumes \cite{She94,Imr95} that the SP state is given as
\begin{equation}
  \psi_k(j)\propto \frac{1}{\sqrt{\lambda_1}} \exp  \left(-
  \frac{|j-k|}{\lambda_1} + i \theta(j) \right)
\label{eq-psi1}
\end{equation}
with $\theta(j)$ a random phase, one finds \cite{She94} that the typical
interaction matrix element has a magnitude of
\begin{equation}
u \propto \lambda_1^{-3/2}
\label{eq-u}
\end{equation}
since it is the sum of $\lambda_1$ random contributions of magnitude
$\lambda_1^{-2}$.  Shepelyansky next calculated the decay rate
$\Gamma$ of a non-interacting eigenstate by means of Fermi's golden
rule $\Gamma \sim U^2/\lambda_1 t$ \cite{She94,She96a,JacS95}. Since
the typical hopping distance is of the order of $\lambda_1$ the
diffusion constant is $D \sim U^2 \lambda_1 /t$.  Within a time $\tau$
the particle pair visits $N \sim U \lambda_1^{3/2} t^{-1/2}
\tau^{1/2}$ states.  Diffusion stops when the level spacing of the
visited states is of the order of the frequency resolution $1/\tau$.
This determines the cut-off time $\tau^*$ and the corresponding
pair-localization length is obtained as $\lambda_2 \sim \sqrt{ D
  \tau^*} \sim (U/t)^2 \lambda_1^2$ in agreement with Eq.\ 
(\ref{eq-lambda2}). Applicability of Fermi's golden rule requires
$\Gamma \gg t/\lambda_1^2$ which is equivalent to $U^2 \lambda_1 /t^2
\gg 1$.  This is exactly the condition for an enhancement of
$\lambda_2$ compared to $\lambda_1$.  Alternatively, the model may be
mapped to a random-matrix model (RMM) with entries chosen according to
Eq.\ (\ref{eq-u}) \cite{She94,FraMP96,FraMP97}.

\section{Numerical results for the random-matrix model of TIP}
\label{sec-tip-rmm}

The arguments presented in the last section are of a qualitatively
nature and Eq.\ (\ref{eq-lambda2}) must be checked for quantitative
accuracy. Even before testing (\ref{eq-lambda2}), it is already
worthwhile to check the validity of (\ref{eq-u}) and the subsequent
arguments or the RMM approach \cite{She94,FraMP96,FraMP97}.  In Ref.\ 
\cite{FraMPW95}, it had been shown that the assumption of a Gaussian
distribution of the matrix elements $u$ was oversimplified.  The
distribution showed long tails making the arithmetic average
unsuitable to characterize the typical value.  In Ref.\ \cite{RomSV99}
we have paid special attention to the exact dependence of $u$ on
$\lambda_1$ and system size.  To this end, we diagonalized the 1D
Anderson model for a given $M$ and $W$ and computed $u$ by averaging
over all suitable states and many disorder configurations.  We showed
that due to the strongly non-Gaussian distribution of $u$, one should
rather use the logarithmic average than the arithmetic average as the
typical value for the computation of $u(\lambda_1)$. But whereas the
arithmetic average \cite{WaiP98} gives $u_{\rm abs}\propto
\lambda_1^{-1.5}$, the typical value obeys $u_{\rm typ}\propto
\lambda_1^{-1.95}$.  Following the arguments above, this would imply
$\lambda_2\propto \lambda_1^{1.1}$, i.e., a very small enhancement. We
emphasize that this result does not mean that there is no enhancement
of the localization length. Rather, the results of Ref.\ 
\cite{RomSV99} indicate that the arguments of Ref.\ \cite{She94}
capture the physics, but only in a somewhat simplified form.  One step
towards a better agreement between the analytical and the numerical
approaches is to take into account the energy denominators in the
computation of $u$, e.g., to consider only interaction matrix elements
for states whose energy spacings are of the order of $U$ or smaller
\cite{RomLS99}.  In this case we find that there is a slight decrease
in the value of the typical exponent and correspondingly a slight
increase in TIP delocalization yielding $\lambda_2 \propto
\lambda_1^{1.4 \pm 0.2}$.  This suggests that higher orders in
perturbation theory than the first order RMM approach \cite{She94} are
important. Furthermore, the exponent ${1.4 \pm 0.2}$ is in reasonable
agreement with previous results in the literature
\cite{FraMPW95,RomS98,Hal96T,HalMK98,HalZMK98,SonO98,WeiP96,SonK97,AkkP97,BriGKT98,HalKK98,WaiP98,WaiWP99,Fra99,LeaRS99,OppWM96,JacS95,JacSS97,Imr95,PonS97}.

\section{RMM approach for toy models}
\label{sec-toy}

In this section we show that a naive application the RMM approach may
give qualitatively incorrect results even if the RMM contains the
correct dependence of the matrix elements on the disorder strength.
To this end we consider two toy models which seem to be closely
related to the TIP problem.  For these models, viz.\ Anderson models
of localization with additional perturbing random potentials, we show
that mapping onto RMMs and estimating the localization length by
Fermi's golden rule leads to an incorrect enhancement of the
localization length.


\subsection{2D Anderson model with perturbation on a line}
\label{sec-2dam}

The first example is set up to lead to the same RMM as the TIP
problem. It consists of the usual 2D Anderson model perturbed by an
additional weak random potential of strength $U$ at the diagonal $x=y$
in real space. Since the perturbation increases the width of the
disorder distribution at the diagonal we expect it to decrease the
localization length.  We map the model onto an RMM following the
arguments for the TIP problem sketched in Sec.
\ref{sec-tip-introduction}.  Again, the eigenstates of the unperturbed
system are localized with a localization length $\lambda_1$ and
approximately given by
\begin{eqnarray}
  \psi_{n} (x,y) \sim {\frac{1}{\lambda_1}} \exp \left[ -\frac {|{\bf r}-{\bf
    r}_n|} {\lambda_1}+ i \theta_n({\bf r}) \right]
\end{eqnarray}
where ${\bf r}= (x,y)^T$ is the coordinate vector of the particle and
$\theta$ is again assumed to be a random phase.  The Hamiltonian of
the 2D perturbed Anderson model differs from the TIP Hamiltonian in
two points: (i) the diagonal elements (given by the random potential)
are independent random numbers instead of being correlated as in the
TIP problem and (ii) the perturbing potential $U(x,x) \in [-U,U]$ at
each diagonal site is random instead of having a definite sign and
modulus $U$ as in the TIP problem. However, none of these points
enters the mapping procedure outlined in Sec.\ 
\ref{sec-tip-introduction}.  Thus, we find that the perturbation
couples each state close to the diagonal ($|x_n - y_n| < \lambda_1$)
to ${\cal O}(\lambda_1^2)$ other such states. The interaction matrix
element is again a sum of ${\cal O} (\lambda_1)$ terms of magnitude
$U/\lambda_1^2$ and random phases and as before $u \sim
U\lambda_1^{-3/2}$.  Consequently, our toy model is mapped onto
exactly the same RMM as TIP in a random potential.  Therefore, the
resulting localization length along the diagonal is also given by Eq.\ 
(\ref{eq-lambda2}).  We thus arrive at the surprising conclusion, that
adding a weak random potential at the diagonal of a 2D Anderson model
leads to an enormous enhancement of the localization length along this
diagonal, in contradiction to the expectation expressed above, viz.\ 
that increasing disorder leads to stronger localization.

\begin{figure}
\centerline{\epsfig{file=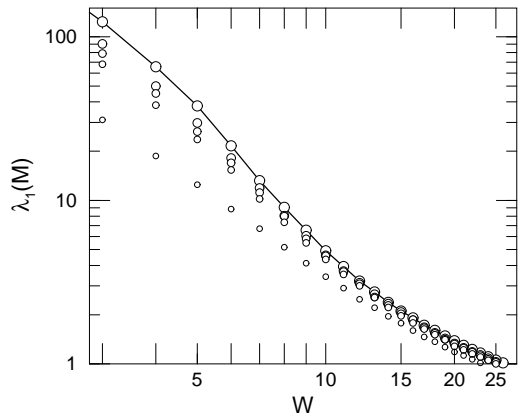,width=\figwidth}}
\vspace{-1cm}
\caption{
  Dependence of $\lambda_1(M)$ on disorder $W$ for the 2D Anderson
  model at $E=0$ for $M= 10, 25, 30, 35$ and $50$ indicated by
  increasing symbol size. We use the $M=50$ data, emphasized by the
  solid line, as finite-size estimate of $\lambda_1$. }
   \label{fig-d2me-l1w}
\end{figure}

As for the TIP case \cite{RomSV99} we now numerically check whether
the relation $u \sim U\lambda_1^{-3/2}$ between the coupling matrix
element $u$ and the localization length $\lambda_1$ of the unperturbed
system is correctly described by the RMM.  Since in 2D a simple
analytic formula for the dependence of $\lambda_1$ on the disorder $W$
does not exist, we first compute estimates $\lambda_1(M)$ for quasi-1D
strips of finite strip width $M$ with $1\%$ accuracy by TMM.  In Fig.\ 
\ref{fig-d2me-l1w}, we show data of $\lambda_1(M)$ as a function of
$W$. In the following, we take $\lambda_1(50)$ to compute the coupling
matrix elements.

Next, we calculate both the arithmetic average $u_{\rm{abs}}= \langle
|u| \rangle$ and the logarithmic average $u_{\rm{typ}}= \exp[\langle
\log(|u|)\rangle]$ for different values of $W$ and various $M \times
M$ squares. Disorder averaging is over $20$ samples and we study
$u_{{\rm abs}}$ and $u_{{\rm typ}}$ as functions of $\lambda_1(M)$.
We emphasize that instead of the well-known extrapolations of
$\lambda_1(M)$ to infinite system size by means of FSS \cite{KraM93},
we take the finite-size approximants $\lambda_1(M)$ on purpose, since
we compute $\lambda_2$ also for comparable finite sizes only.

\begin{figure}[t]
\centerline{\epsfig{file=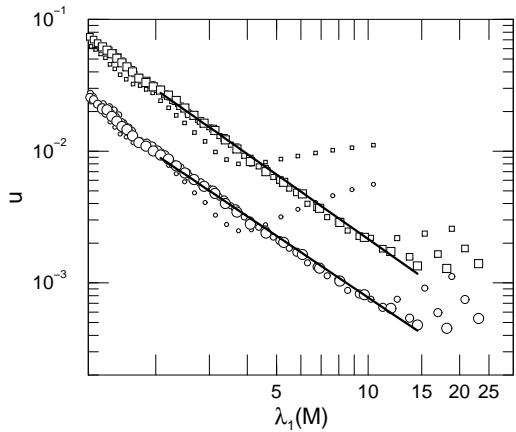,width=\figwidth}}
\vspace{-1cm}
\caption{
  Dependence of $u_{{\rm abs}}$ (squares) and $u_{{\rm typ}}$
  (circles) on $\lambda_1(M)$ for the perturbed 2D Anderson model with
  $U=1$ and $M= 10, 25, 30$ and $35$ indicated by increasing symbol
  size. The solid lines represent the power laws $u_{{\rm abs}} \sim
  \lambda_1^{-1.6}$ and $u_{{\rm typ}} \sim \lambda_1^{-1.5}$. }
   \label{fig-d2me-l1}
\end{figure}

As for the TIP model \cite{RomSV99} the distribution $P_{{\rm o}}(u)$
of the (off-diagonal) coupling matrix elements is strongly
non-Gaussian, suggesting that $u_{{\rm typ}}$ rather than $u_{{\rm
    abs}}$ is the relevant quantity.  The results for $u_{{\rm abs}}$
and $u_{{\rm typ}}$ are presented in Fig.\ \ref{fig-d2me-l1}.
The dependence of $u_{{\rm abs}}$ on $\lambda_1(M)$ for $2 \leq
\lambda_1(M) \leq 12$ follows $u_{{\rm abs}} \propto
\lambda_1(M)^{-1.6 \pm 0.1}$ in agreement with the RMM value of 3/2
and with Ref.\ \cite{RomSV99}.  Furthermore, here we also have
$u_{{\rm typ}} \propto \lambda_1(M)^{-1.5 \pm 0.1}$.  We note that the
change of the slopes of $u_{{\rm abs}}$ and $u_{{\rm typ}}$ at
$\lambda_1(M)\approx M/2$ is entirely due to the finite sample sizes
\cite{RomSV99}.

Consequently, in contrast to the TIP problem the RMM model for the 2D
perturbed Anderson model of localization contains the correct
dependence of the coupling matrix elements on the localization length
of the unperturbed system, but still it leads to an incorrect
enhancement of the localization length along the diagonal.


\subsection{1D Anderson model with perturbation}
\label{sec-1dam}

An even more striking contradiction can be obtained for an 1D Anderson
model of localization. The eigenstates are again given by Eq.\ 
(\ref{eq-psi1}) with $\lambda_1$ known from second order perturbation
theory \cite{Eco90,KapW81,Tho79} and numerical calculations
\cite{CzyKM81,Pic86} to vary as $\lambda_1 \sim t^2/W^2$ for small
disorder. We now add a weak random potential of strength $U$ at all
sites. Since the result is obviously an 1D Anderson model with a
slightly higher disorder strength $W_u > W$ the localization length
will be reduced, $\lambda(U) \sim t^2/W_{u}^{2}$.

The mapping onto an RMM can be performed in complete analogy to the
TIP problem and the 2D Anderson model discussed above.  The perturbing
potential leads to transitions between the unperturbed eigenstates
$\psi_n$.  Each such state is now coupled to ${\cal O}(\lambda_1)$
other states by coupling matrix elements $\langle \psi_{n}|U|\psi_{n'}
\rangle$ with magnitude $u \sim U \lambda_1^{-1/2}$ since we sum over
$\lambda_1$ contributions with magnitude $U/\lambda_1$ and supposedly
random phases.

The application of Fermi's golden rule in this 1D case leads to a
diffusion constant $D \sim U^2 \lambda_1^2 /t$. The number of states
visited within a time $\tau$ is $N \sim U \lambda_1 t^{-1/2}
\tau^{1/2}$.  Again, diffusion stops at a time $\tau^*$ when the level
spacing of the states visited equals the frequency resolution.  This
gives $\tau^* \sim U^2 \lambda_1^2 /t^3$.  The localization length
$\lambda$ of the perturbed system thus reads $\lambda \sim \sqrt{D
  \tau^*} \sim U^2 \lambda_1^2$ as in Eq.\ (\ref{eq-lambda2}), in
contradiction to the correct result.

\begin{figure}
\centerline{\epsfig{file=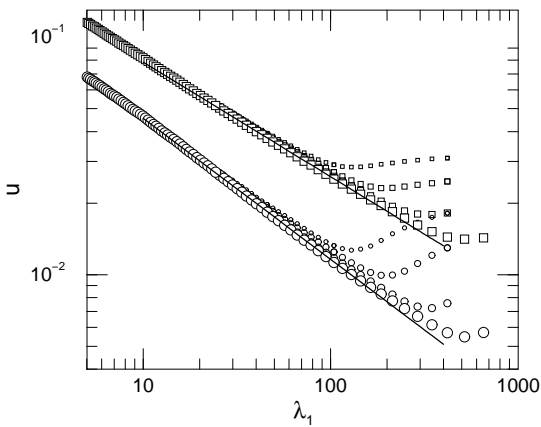,width=\figwidtha}}
\vspace{-1cm}
\caption{
  Dependence of $u_{{\rm abs}}$ (squares) and $u_{{\rm typ}}$
  (circles) on $\lambda_1$ for the perturbed 1D Anderson model with
  $U=1$ and $M= 200, 300, 500$ and $800$ indicated by increasing
  symbol size.  The solid lines represent the power laws $u_{{\rm
      abs}} \sim \lambda_1^{-0.48}$ and $u_{{\rm typ}} \sim
  \lambda_1^{-0.59}$. }
   \label{fig-d1me-l1}
\end{figure}

Again we numerically check the relation between $u_{{\rm abs}}$ and
$u_{{\rm typ}}$ and the unperturbed localization length $\lambda_1$.
In Fig.\ \ref{fig-d1me-l1}, we show results obtained for chains with
various lengths and $50$ disorder configurations for each $W$.
$\lambda_1$ is computed by TMM.
For $10 \leq \lambda_1 \leq 250$, $u_{{\rm abs}}$ varies as
$\lambda_1^{-0.48 \pm 0.10}$ as predicted above. $u_{{\rm typ}}$
varies as $\lambda_1^{-0.59 \pm 0.10}$. Both variations are compatible
with the RMM value of $1/2$ for the exponent.  Again we need
$\lambda_1 < M/2$ in order to suppress finite size effects.

Consequently, although the RMM model for the 1D perturbed Anderson
model of localization contains the correct dependence of the coupling
matrix elements on the localization length of the unperturbed system,
it still leads to an incorrect enhancement of the localization length.

\section{Application of the block-scaling picture to toy models}
\label{sec-bsptoy}

Let us now discuss the relation of these results to Imry's
block-scaling picture (BSP) \cite{Imr96,Imr95} for the TIP problem.
In this approach one considers blocks of linear size $\lambda_1$ and
calculates the dimensionless pair conductance on that scale,
\begin{equation}
  g_2 \sim \frac{u^2}{\Delta^2},
  \label{eq-bsp}
\end{equation}
where $u$ represents the typical interaction-induced coupling matrix
element between states in neighboring blocks and $\Delta \sim
t/\lambda_1^2$ is the level spacing within the block. If the typical
coupling matrix element depends on $\lambda_1$ as $u \sim U
\lambda_1^{-\alpha}$ the pair conductance obeys
\begin{equation}
  g_2 \sim  (U/t)^2 \lambda_1^{4-2\alpha}.
\end{equation}
For the 2D Anderson model with perturbation considered above, the BSP
can be applied analogously. Again, we consider blocks of linear size
$\lambda_1$ and compute the typical perturbation-induced matrix
elements between these blocks.  We then find that according to the BSP
the conductance of a 2D Anderson model with additional weak perturbing
potential along the diagonal is given by Eq.\ (\ref{eq-bsp}). Using
$\alpha= 1.5 \pm 0.1$ as obtained above from the numerical data for
$u_{{\rm abs}}$ and $u_{{\rm typ}}$, we then have $g_2 \sim (U/t)^2
\lambda_1$. Thus we conclude that the BSP does not work for our 2D toy
model, because it yields the same unphysical result as the RMM
approach of section \ref{sec-2dam}.

Let us also apply the BSP to the 1D toy example. The level spacing in
a 1D block of size $\lambda_1$ is $\Delta \sim t/\lambda_1$, and the
coupling matrix element between states in neighboring blocks is $u
\sim U\lambda_1^{-1/2}$.  Thus, the conductance of the perturbed
system on a scale $\lambda_1$ is obtained as $ g_2 \sim (U/t)^2
\lambda_1$.  For large $\lambda_1$ this again contradicts the correct
result, viz.\ a decrease of the conductance compared to the
unperturbed system.  Thus, the BSP applied to the two toy models
introduced in the present work gives the same qualitatively incorrect
results for the localization properties as the RMM. This is not
surprising since the only ingredients of the BSP are the intra-block
level spacing $\Delta \sim t/\lambda_1^2$ and the inter-block coupling
matrix elements $u$ which also enter the RMM.

\section{Conclusions}
\label{sec-concl}

We have presented two toy models which seem to be closely related to
the TIP problem. For these toy models the usual analytical arguments
given to support the delocalization of TIP, viz.\ the RMM and the BSP
do not work. However, the large-scale numerical simulations
\cite{SonO98,SonK97,Fra99,LeaRS99,RomLS99b,OrtC99} have convincingly
shown that an enhancement of the two-particle localization length due
to the interaction exists, even though the detailed results are more
complicated than the original prediction (\ref{eq-lambda2}). This
leads, of course, to the question, under which conditions the RMM
mapping and the BSP give the correct result, at least qualitatively.
While a general answer to this question is not known, it has been
suggested \cite{Wei98} that the difference between the TIP and our toy
models is an additional symmetry in the TIP problem.

In summary, the two examples suggest that additional physical insight
is needed before applying the RMM. In addition, we expect that taking
into account the energies of the states as in Ref.\ \cite{RomLS99} for
TIP will result in a reduced enhancement, i.e., a smaller value of
$\alpha$, in the analytical predictions. This will in turn give a better
agreement with the numerically determined dependence of the TIP
localization length on $\lambda_1$.


\end{document}